\documentstyle[12pt,dina4p]{article}
\newcommand{\nc}{\newcommand}
\nc{\rnc}{\renewcommand}
\nc{\bm}{\bibitem}
\def\3{\ss}
\nc{\mb}{\mbox}
\nc{\be}{\begin{equation}}
\nc{\ee}{\end{equation}}
\nc{\bea}{\begin{eqnarray*}}
\nc{\beam}{\begin{eqnarray}}
\nc{\eea}{\end{eqnarray*}}
\nc{\eeam}{\end{eqnarray}}
\nc{\nn}{\nonumber}
\nc{\dps}{\displaystyle}
\nc{\nichts}{\rule{0ex}{1ex}}
\nc{\wenig}{\hspace*{1cm}}
\nc{\etwas}{\hspace*{2cm}}
\nc{\mehr}{\hspace*{3cm}}
\nc{\Rz}{Raum-Zeit}
\nc{\ped}{positive-Energie-Darstellung}
\nc{\hkn}{Haag-Kast\-ler-Netz}
\nc{\nel}{Netze auf dem eindimensionalen Lichtkegel}
\nc{\npf}{$n$-Punkt-Funktion}
\nc{\zpf}{2-Punkt-Funktion}
\nc{\cst}{$C^{\ast}\!$-Al\-ge\-bra}
\nc{\csts}{$C^{\ast}\!$-Al\-ge\-bren}
\nc{\fO}[1]{\mb{${\cal#1(O)}$}}
\nc{\NO}[1]{\mb{${\cal A:=\overline{\bigcup_{O}#1(O)}^{\|\cdot\|}}$}}
\nc{\fOO}[1]{\mb{${\cal #1 : O \rightarrow #1(O)} $}}
\nc{\fOOO}[1]{\mb{${\cal #1}: I \rightarrow{\cal #1}(I)$\,, $I\in\K$}}
\nc{\FOOO}[1]{\mb{${\cal #1}: I \rightarrow{\cal #1}(I)$\,, $I\in\KK$}}
\nc{\A}{\mb{${\cal A}$}}
\nc{\B}{\mb{${\cal B}$}}
\nc{\BH}{\mb{$B(H)$}}
\nc{\C}{\mb{\bf C}}
\nc{\CC}{\mb{${\cal C}$}}
\nc{\D}{\mb{${\cal D}$}}
\rnc{\d}{\mb{d}}
\nc{\dd}{\!\mb{{\footnotesize {\it d}}}}
\nc{\e}{\mb{$\varepsilon$}}
\nc{\F}{\mb{${\bf{\cal F}}$}}
\nc{\fn}{\mb{$F_n$}}
\nc{\Fn}{\mb{$\overline{F}_n$}}
\nc{\G}{{\it G}}
\rnc{\H}{{\it H}}
\nc{\HH}{\mb{{\bf H}}}
\nc{\k}{\mb{${\cal K}_0$}}
\nc{\K}{\mb{${\cal K}$}}
\nc{\KK}{\mb{$\overline{\K}$}}
\nc{\KKK}{\mb{\scriptsize ${\cal K}_0$}}
\rnc{\l}{\mb{$\lambda$}}
\nc{\N}{\mb{${\bf N}$}}
\rnc{\o}{\mb{$\omega$}}
\nc{\OM}{\mb{$\Omega$}}
\rnc{\O}{\mb{${\cal O}$}}
\rnc{\P}{\mb{{\bf P}}}
\nc{\R}{\mb{\bf R}}
\nc{\Rp}{\mb{$\R_{+}$}}
\nc{\Rm}{\mb{$\R_{-}$}}
\nc{\r}{\mb{\scriptsize \bf R}}
\nc{\rp}{\mb{$\r_{\mb{\tiny +}}$}}
\nc{\rmm}{\mb{$\r_-$}}
\rnc{\S}{\mb{${\cal S}$}}
\nc{\SL}{\mb{$SL(2,\R)/\Z_{2}$}}
\nc{\SO}{\mb{${\cal S(O)}$}}
\nc{\SpP}{\mb{Sp\,{\bf P}}}
\nc{\SR}{\mb{${\cal S(\R)}$}}
\rnc{\t}{\mb{$\vartheta$}}
\nc{\U}{{\it U}}
\rnc{\v}{\mb{$\varphi$}}
\nc{\vg}{\overline{\v_g}}
\nc{\Z}{\mb{\bf Z}}
\title{The Construction of Pointlike Localized
                 Charged Fields
from Conformal Haag-Kastler Nets
\thanks{hep-th/9506016}
}
\author{
Martin J\"or\ss\ \thanks{e-mail address: I02joe@dsyibm.desy.de}
\\
II. Institut f\"ur Theoretische Physik der
Universit\"at Hamburg }
\date{\nichts}
\begin{document}
\thispagestyle{empty}
\maketitle
\begin{abstract}
Starting from a chiral conformal Haag-Kastler net on 2 dimensional
Minkowski space we construct associated charged
                                        pointlike localized fields
which intertwine between
    arbitrary       superselection sectors with finite statistics
                                                of the theory.
This amounts to a proof of the Spin-Statistics theorem,
the PCT theorem and
                                                        the existence
                           of operator product expansions.

This paper generalizes similar results of a recently published paper
by Fredenhagen and the author
                   \cite{FrJ} from the neutral
                                              vacuum sector
to the
the full theory with arbitrary charge and finite statistics.
\end{abstract}
\section{Introduction}
The formulation of quantum field theory can be done
                                 in terms of
Haag Kastler nets of local observable algebras
(``local quantum physics" \cite{Haa}) or
in the framework of
                                          Wightman axioms \cite{StW}.
Usually, both concepts are assumed to be physically equivalent.
There is, however, still no proof for this assumption.

In a recently published paper \cite{FrJ} Fredenhagen and the author
could give a partial answer to the question how both concepts are
interrelated in the case of chiral conformal quantum field theory.
Starting from a chiral conformal Haag-Kastler net on 2 dimensional
Minkowski space we constructed associated neutral
                                        pointlike localized fields
in the vacuum sector.
This amounted to a proof  of
                                                        the existence
                      of operator product expansions for observables.
The present paper generalizes these results
                   from the neutral
                                              vacuum sector
to the
the full theory with arbitrary charge and finite statistics.

                           The existence of
operator product expansions
in the Wightman framework
                            has been postulated by Wilson
                            \cite{Wil}.
                         Especially in 2d conformal field theory,
this assumption has turned out to be very fruitful.
The existence of a convergent expansion of the
product of two fields on the vacuum could been derived
from conformal covariance, but the existence of the associated
local fields had to be postulated \cite{Lue,Mac,SSV}.

                              The existence of an
operator product expansion
in the Haag-Kastler framework
                           might be formulated as the existence of
sufficiently many Wightman fields such that their linear span
applied to the vacuum is dense in the Hilbert space.
Actually, we are able to derive a stronger result:
    We prove an expansion of local operators (arbitrary local
                                                            elements
of the reduced field bundle     \cite{FRS1,FRS2}
                            associated to the net of observables)
                                                               into
  charged                                                     fields
with local coefficients and show that this expansion
converges $*$-strongly on a dense domain in the physical
                                             Hilbert space.

The construction of Wightman fields out
of Haag-Kastler nets should be possible
                     by some scaling limit.
 This heuristic idea, however,             is
difficult to formulate in an intrinsic way \cite{Buc}.
Buchholz and Fredenhagen \cite{BuF} performed
                                          the construction
of a pointlike field
in the presence of massless particles
                                           in a dilation invariant
theory where scaling is well defined.

We study the possibly simplest
situation: Haag-Kastler nets in 2 dimensional Minkowski
space with trivial translations in one light cone
direction (``chirality") and covariant under
the real M\"obius group which acts on the other lightlike
direction.

We construct pointlike localized fields carrying
arbitrary charge with finite statistics
                 and therefore intertwining between the different

superselection sectors of the theory.
(In Conformal Field Theory
 these objects are known as ``Vertex Operators".)
                                      We obtain
                                          the unbounded field
operators as limits of elements of the reduced field bundle
                             \cite{FRS1,FRS2}
                                                            associated
to the net of observables of the theory.
                                               Their
smeared linear combinations are affiliated to the reduced field bundle
 and generate it. At the moment, we do not know whether
the constructed fields satisfy all Wightman axioms,
since we have not yet found
      an invariant domain of definition.

Our method consists of an explicit use of the
representation theory of the universal covering group of
                         \mb{$SL(2,\R)$} combined with
a conformal cluster theorem. This conformal cluster theorem
                            was proven in \cite{FrJ} for the vacuum
sector and can
be generalized in the present paper to the case of arbitrary charge
with finite statistics.

As a consequence of the existence of charged pointlike localized fields
we can finally prove
                    the Spin-Statistics theorem\footnote{After
                       the completion of the present paper we
received a preprint by Guido and Longo \cite{GLo} that gives an
independent proof of the conformal Spin-Statistics theorem.},
                                                the PCT theorem for the
full theory,
               a generalized version of the Bisognano-Wichmann property
and additivity of the net of local algebras.

\section{Assumptions and Results}
\subsection{Assumptions}
Let $\A=(\A(I))_{I\in\KKK}$
                       be a family of von Neumann algebras on some
separable Hilbert space \H. \k\ denotes the set of nonempty bounded
open intervals on \R.
$\A$ is assumed to satisfy the following conditions.
\begin{enumerate}
\def\labelenumi{\roman{enumi})}
\def\theenumi{\roman{enumi}}
\item Isotony:
\be \A(I_1)\subset\A(I_2)\;\;\;\;\;\mb{for}\;\;\;\;I_1\subset I_2,
\;\;\;\;I_1, I_2\in\k.\ee
\item Locality:
\be \A(I_1)\subset\A(I_2)'\;\;\;\;\;\mb{for}\;\;\;\;I_1\cap I_2=\{\},
\;\;\;\;I_1, I_2\in\k\ee
($\A(I_2)'$ is the commutant of $\A(I_2)$).
\item
There exists a strongly continuous unitary representation $U$ of
$G=SL(2,\R)$ in $H$ with $U(-1)=1$ and
\be U(g)\,\A(I)\,U(g)^{-1}=\A(gI),\;\;\;\;\;I,gI\in\k
\ee
($SL(2,\R)\ni g=\left(\begin{array}{cc}a&b\\c&d\end{array}\right)$
                       acts
                            on $\R\cup\{\infty\}$ by
$x\mapsto
\frac{ax+b}{cx+d}$ with the appropriate interpretation for
$x, gx=\infty$).
\item
The conformal Hamiltonian \HH, which generates the
    restriction of $U$ to $SO(2)$, has nonnegative spectrum.
\item
There is a unique (up to a phase) $U$-invariant unit vector
$\OM\in\H$.
\item
\H\ is the smallest closed subspace containing the vacuum
                                               $\OM$ which
is invariant under $U(g),$ $g\in SL(2,\R),$ and $A\in\A(I),I\in\k$
(``cyclicity").
\end{enumerate}

\subsection{Known Results}
Several results on the structure of the vacuum sector of an
conformally invariant Haag-Kastler theory in 1+1 dimensions are
well known.
 With the use of
          a theorem of Borchers \cite{Bor} it was possible
  \cite{FrG,BGL}                                          to
identify the modular structure with geometrical objects
in accordance with the ideas of Bisognano and Wichmann
    \cite{BiW}.
As a consequence, Haag duality and additivity of the net of
 local algebras and a PCT theorem could be proven.
In \cite{FrJ}, we recently constructed pointlike localizable
conformal field operators in the vacuum sector.
                                 These unbounded operators
of neutral fields could be shown to be closable, their closures
are affiliated with the associated algebra of local observables
and generate the local net of observables.

In the same paper,
a conformal cluster theorem could be proven and an expansion
of local observables into neutral fields with local
coefficients was given.

\subsection{New Results}
In order to construct charged fields intertwining between the
superselection sectors with finite statistics
                       of the theory, we consider
the reduced field bundle $\F_{red}=(\F_{red}(I))_{I\in\KKK}$ associated
to the net of observables $\A=(\A(I))_{I\in\KKK}$, introduced in
\cite{FRS1,FRS2}.

                  The reduced field bundle $\F_{red}$ is an algebra
densely spanned by operators $F=F(e,A)$, linear in the local degree
of freedom $A\in\A$ and with a multi-index $e$. This multi-index
                                               refers to the charge
carried by $F$ as well as to the source sector
and the range sector between which $F$ interpolates according
to the ``fusion rules". If $\rho,\alpha,\beta$ are charge, source
and range of the corresponding ``fusion channel" the multi-index
$e$ is said to be of type $(\alpha,\rho,\beta)$.
                  The elements of the reduced field bundle act on
                           $\H_{red}$, a realization of
the physical Hilbert space.
$\H_{red}$ is the direct sum of copies of the vacuum Hilbert space,
one for each superselection sector with finite statistics.
The direct sum of the representations of the universal covering
group of the conformal group for each superselection sector with finite
statistics                                                  will
be called $U(\tilde{G})$.

We are able to prove a generalization of the
conformal cluster theorem to arbitrary charged sectors with finite
statistics.                                            This theorem
specifies the decrease properties of conformal
two-point-functions in the algebraic framework to be exactly
those known from theories with pointlike localization.

\medskip

{\bf Theorem:}
Let $(\A(I))_{I\in\KKK}$ be a conformally covariant local net on $\R$.
Let       $a,b,c,d\in\R$ and
              $a<b<c<d$.
Let $F\in\F_{red}(\,(a,b)\,)$, $G\in\F_{red}(\,(c,d)\,)$, $n\in\R$ and
$P_k F\OM=P_k \bar{F}\OM=0,\;k<n$. $P_k$ here denotes the
projection on the subrepresentation of $U(\tilde{g})$
                                        with conformal dimension $k$.
We then have
\be
|(\,\OM,GF\OM\,)|\leq \left(\frac{(b-a)\,(d-c)}{(c-a)\,(d-b)}\right)^n
\;\|F\|\,\|G\|.
\ee

\medskip

Due to the positivity condition, the representation
              $U(\tilde{G})$ is completely
reducible into irreducible subrepresentations
         and the irreducible components $\tau$ are up to equivalence
uniquely characterized by the conformal dimension $n_\tau\in\Rp$
($n_\tau$ is the lower bound of the spectrum of the conformal
 Hamiltonian \HH\ in the representation $\tau$).

Associated with each irreducible subrepresentation $\tau$ of $U$
we find
for each $I\in\k$ a
     densely defined operator valued distribution $\varphi_\tau^I$
on the space $\D(I)$
             of Schwartz functions with support in $I$
   such that the following statements hold for all $f\in\D(I).$
\begin{enumerate}
\def\labelenumi{\roman{enumi})}
\def\theenumi{\roman{enumi}}
\item
The domain of definition of
$\varphi_\tau^I(f)$ is given by $\A(I')\,\OM$.
\item
\be
\varphi_\tau^I(f)\,\OM\in P_\tau\H_{red}
\ee
with $P_\tau$ denoting the projector on the module of $\tau$.
\item
\be
U(\tilde{g})\;
    \varphi_\tau^I(x)\;
        U(\tilde{g})^{-1}=(cx+d)^{-2n_\tau}\varphi_\tau^{gI}
                                 (\tilde{g}x)
\ee
with the covering projection $\tilde{g}\mapsto g$ and
$g=\left(\begin{array}{cc}a&b\\c&d\end{array}\right)\in
SL(2,\R),\;I, gI\in\k$.
\item
$\varphi_\tau^I(f)$ is closable.
\item The exchange algebra of the reduced field bundle \cite{FRS2}
and the existence of the closed field operators
$\varphi_\tau^I(f)$, mapping a dense set of the vacuum Hilbert space
into some charged sector with finite statistics,
                          suffice to construct  closed field operators
$\varphi_{\tau,\alpha}^I(f)$,
                             mapping a dense set of an arbitrary charged
sector $\alpha$ with finite statistics
                into
                     some (other) charged sector with finite statistics.
Here,
                    the irreducible
                                          module $\tau$ of
$U(\tilde{G})$ labels orthogonal irreducible fields defined
in the same sector $\alpha$\,.
\item
The closure of any $\varphi_{\tau,\alpha}^I(f)$ is affiliated to
                                          $\F_{red}(I)$.
\item
$\F_{red}(I)$ is the smallest von Neumann algebra to which all operators
$\varphi_{\tau,\alpha}^I(f)$ are affiliated.
\end{enumerate}

With the existence of pointlike localized fields we are able to
give a proof of a generalized Bisognano-Wichmann property. We can
identify the conformal group and the reflections as generalized
modular structures in the reduced field bundle. Especially, we obtain
a PCT operator on $\H_{red}$ proving the PCT theorem for the full
theory.

Moreover, the existence of pointlike localized fields gives a
proof of the hitherto unproven Spin-Statistics theorem for conformal
Haag-Kastler nets in 1+1 dimensions.

Its also possible to generalize the operator product expansions
of \cite{FrJ} to charged sectors with finite statistics:
\medskip

{\bf Theorem:} For each $I\in\k$ and each $F\in \F_{red}(I)$
there is a local expansion
\be
F\,=\,\sum_{\tau,\alpha}\varphi_{\tau,\alpha}^I(f_{\tau,\alpha,F})\,
\ee
into a sum over all sectors $\alpha$ with finite statistics
                                     and all irreducible modules
$\tau$ of $U(\tilde{G})$
with \be
 \mb{supp}f_{\tau,\alpha,F}\subset I\,,\ee
which converges on $\F_{red}(I')\OM$ $*$-strongly (cf. the definition in
\cite{BrR}). Here, $I'$ denotes the complement of $I$ in $\k$.

\medskip
\section{Construction of Pointlike Localized Charged Fields}
The following
    idea for the definition of charged
                               conformal fields is the
generalization of the idea for the case of neutral fields:
Let $A$ be a local observable,
$A\in\A(I_0)$, $I_0\in\k,$
let $\rho$ be a localized and transportable endomorphism of $\A$
inducing a charged sector with finite statistics,
                                let $e$ be a field bundle
                                        multi-index of type
$(0,\rho,\rho)$. Then $F=F(e,A)$ is a local element of the
reduced field bundle. Let now $n\in\Rp$
and $P_n$ the projection on the subspace of conformal dimension
$n$ in $H_{red}$.
    We can think of $P_n F\OM$  as a vector of the form
$\varphi_{n}(h)\,\OM$ where $\varphi_{n}$ is a conformal field of
dimension $n$ and $h$ is an appropriate
                                function on $\R$.
                                           The exact relation between
$F$ and $h$, however, is unknown
at the moment.
                        All we have are the
transformation properties under $\tilde{G}$:
\be
U(\tilde{g})\,P_n F\OM=\varphi_{n}(h_{\tilde{g}}^{(n)})\,\OM
\ee
with $h_{\tilde{g}}^{(n)}(x)=(cx-a)^{2n-2}\,h(\frac{dx-b}{-cx+a})$
and the covering projection $\tilde{g}\mapsto g$ for
$g=\left(\begin{array}{cc}a&b\\c&d\end{array}\right)\in G$.
Scaling the vector $P_n F\OM$ by dilations $D(\l)=
U\left(\begin{array}{cc}\l^{\frac{1}{2}}&0\\0&\l^{-\frac{1}{2}}
\end{array}\right)$ we find
\be
D(\l)\,P_n F\OM=\l^n\,\varphi_{n}(h_{\l})\,\OM
\ee
with $h_{\l}(x)=\l^{-1}\,h(\frac{x}{\lambda
                                     })$. Hence, we obtain formally
for $\lambda\downarrow 0$
\be
\l^{-n}\,D(\l)\,P_n F\OM\longrightarrow
                                          \int dx\,h(x)\;\varphi_{n}
                                                                     (0)
\,\OM.
\ee
                                                        We smear over
the group of
    translations $T(a)=U\left(\begin{array}{cc}1&a\\0&1\end{array}
\right)$ with some test function $f$ and obtain in the formal limit
a Hilbert space vector:
\be
\label{a}
\lim_{\lambda
        \downarrow 0}\l^{-n}\int da\,f(a)\;T(a)\,D(\l)\,P_n F\OM=
\int dx\,h(x)\;\varphi_{n}(f)\,\OM.\label{f}
\ee
Now, the left hand side can be interpreted
                        as a definition of a conformal
field $\varphi_{n}$ on the vacuum vector $\OM$.
Writing down
\be
\varphi_{n}^I(f)\,A'\OM=\pi_\rho(A') \varphi_{n}^I(f)\,\OM,\;\;
f\in\D(I),\,A'\in\A(I'),\,I\in\k,
\ee
we obtain operators with a domain of definition that is dense in
the vacuum Hilbert space. In this section, $I'$ always
                                                denotes the complement
of $I$ in $\k$.

In order to make this formal construction meaningful
there are two problems to overcome.

The first one is the fact that the limit on the left hand side of
(\ref{a})
     does not exist in general if $F\OM$ is replaced by an arbitrary
vector in $H_{red}$.
                 This corresponds to the possibility that the function
$h$ on the right hand side might not be integrable.
After smearing the operator $F$
                                          with a smooth function on
                                                         $\tilde{G}$
the limit is well defined. Such operators will be called regularized.

The second problem is to show that the smeared field operators
$\varphi_{n}^I(f)$ are closable, in spite of the nonlocal
nature of the projections $P_n$. This problem will be solved by
an argument
based on the generalization of the
                              conformal cluster theorem \cite{FrJ}
to the charged case.

First, we generalize the conformal cluster theorem to charged sectors
with finite statistics.

\medskip

{\bf Theorem:}
Let $(\A(I))_{I\in\KKK}$ be a conformally covariant local net on $\R$.
Let       $a,b,c,d\in\R$ and
              $a<b<c<d$.
Let $F\in\F_{red}(\,(a,b)\,)$, $G\in\F_{red}(\,(c,d)\,)$, $n\in\R$ and
$P_k F\OM=P_k \bar{F}\OM=0,\;k<n$. $P_k$ here denotes the
projection on the subrepresentation of $U(\tilde{g})$
                                        with conformal dimension $k$.
We then have
\be
|(\,\OM,GF\OM\,)|\leq \left(\frac{(b-a)\,(d-c)}{(c-a)\,(d-b)}\right)^n
\;\|F\|\,\|G\|.
\ee

\medskip

{\bf Proof:} (Cf.\,the proof of the conformal cluster theorem
in \cite{FrJ} and the idea in \cite{Fre}\,)

             Choose $R>0$. We consider the following 1-parameter
subgroup of $G=SL(2,\R)$
\be
g_t\,:\,x\longmapsto\frac{x\,
                       \mb{cos}\frac{t}{2}+R\,\mb{sin}\frac{t}{2}}
{-\frac{x}{R}\,\mb{sin}\frac{t}{2}+\mb{cos}\frac{t}{2}}\,.
\ee
Its generator ${\rm \bf H}_R$ is within each subrepresentation
of $U(\tilde{G})$
                                 unitarily equivalent to the
conformal Hamiltonian ${\rm \bf H}$. Therefore, the spectrum of $F\OM$
and $\bar{F}\OM$ w.r.t.\ ${\rm \bf H}_R$ is bounded below by $n$.
Let $-\pi<t_0<t_1<\pi$ such that $g_{t_0}(b)=c$ and $g_{t_1}(a)=d$.
Because of the conformal covariance of the reduced field bundle,
the function
\be
M(z):=
(\,\OM,\,G\,\alpha_{g_t}(F)\,\OM\,)\,,\;\;z=e^{it},\,-\pi<t<\pi,\,
                                                     t\not\in
[t_0,t_1]\,,
\ee
is well-defined in its domain of definition.
We consider the analytical properties of
\be
N(z):=\,(z-z_0)^n\,(z^{-1}-z_0^{-1})^n\,M(z)\,,\;\;
                         z_0:=e^{\frac{i}{2}(t_0+t_1)}\,.
\ee
     Using the condition of positive energy
and weak locality
           $N(z)$ can easily be continued analytically
                                         (see \cite{FrJ}).
We find singularities at
              $z=0$, $z=\infty$ and on (the copies of) the
                                           interval
$[e^{it_0},e^{it_1}]$
and branch-cuts (with arbitrary position) have to be introduced
                                 which  connect the singularities.
Hence, as maximal domain of analyticity we obtain a Riemannian
surface.
To apply the maximum principle of complex analysis
we consider the continuation of $N(\cdot)$ to the
     Alexandroff-compactification of this
Riemannian surface at $z=0$ and $z=\infty$
                   (see \cite{Str}).
In vicinities of
$z=0$ and $z=\infty$ the function
                   $N(\cdot)$ is bounded
because of the bound on the spectrum of ${\rm \bf H}_R$ and
can therefore be continued analytically to the compactification
                                        \cite{For}.
As an analytic function on the compactified Riemannian surface
                           it reaches its maximum on the boundary of its
domain of definition, i.e.\,on (the copies of)
$[e^{it_0},e^{it_1}]$\,.
                                             Therefore, we obtain
the bound:
\be
\mb{sup}|N(\cdot)|\,\leq\,\|F\|\,\|G\|\,|e^{it_0}-e^{\frac{i}{2}
(t_0+t_1)}|^{2n}\,=\,\|F\|\,\|G\|\,(2\,
                                     \mb{sin}\frac{t_0-t_1}{4})^{2n}
\,.
\ee
This leads, as in \cite{FrJ}, to
\beam
|(\,\OM,\,GF\OM\,)|&=&|M(1)|\,=\,|N(1)|\ |1-e^{\frac{i}{2}(t_0+t_1)}
                                                              |^{
-2n}\,=\,|N(1)|\ (2\,\mb{sin}\frac{t_0+t_1}{4})^{-2n}\nn\\
&\leq&\mb{sup}|N|\,(2\,\mb{sin}\frac{t_0+t_1}{4})^{-2n}\,\leq\,
\|F\|\,\|G\|\,\left(\frac{\mb{sin}\frac{t_0-t_1}{4}}{\mb{sin}
\frac{t_0+t_1}{4}}\right)^{2n}\,.
\eeam
Finally, $t_0$ and $t_1$ remain to be determined. We obtain
\be
\lim_{R\rightarrow\infty}R\,t_0=2(c-b)\;\;\;\mb{and}\;\;\;
\lim_{R\rightarrow\infty}R\,t_1=2(d-a)\,.
\ee
We assume $a-b=c-d$ and find $\left(\frac{t_0-t_1}{t_0+t_1}
\right)^2=\frac{(a-b)\,(c-d)}{(a-c)\,(b-d)}=:x\,.$
The bound on $|(\,\OM,\,GF\OM\,)|$ can only depend
on the conformal cross ratio $x$\,.
Hence, we can drop the assumption and the theorem is
proven.

\subsection{Existence of the Pointlike Field Limit in Charged Sectors}

$P_n H$ can be identified with copies of
                                  $L^2(\Rp,p^{2n -1}\,dp)$,
where $G$ acts according to
\be
\left(U_n\widetilde{
         \left(\begin{array}{cc}a&b\\c&d\end{array}\right)}\Phi
\right)(p)=\lim_{\e\downarrow 0}\frac{1}{2 \pi}\int_{\r} dx\int_{\rp}
dq\,e^{-ip(x+i\varepsilon)
              +iqg^{-1}(x+i\varepsilon)}
                            (a-c(x+i\e))^{2n-2}\Phi(q).
\ee
This fact can be used to
            investigate the limit in
                                 (\ref{a}):
Let $\Phi\in P_n H$ now be smeared out with a test function on
$\tilde{G}$ such that $\Phi$ is $\CC^\infty$, i.e.\
                             $\tilde{g}\mapsto U_n(\tilde{g}
                                                   )\,\Phi$
is $\CC^\infty$. In the appendix of \cite{FrJ} we have proven
in the vacuum case  that
                  smeared out
functions
                                         $\Phi(\cdot)$
                                                are continuous and
bounded in $p$. The argument in \cite{FrJ} uses an expansion into
normalized associated Laguerre polynomials and can be fully transferred
to charged sectors. Having proven continuity and boundedness,
                straight forward calculation leads to
\be
\left(\int da\,f(a)\,T(a)\,D(\l)\,\l^{-n}\,\Phi\right)(p)=
\tilde{f}(p)\,\Phi(\l p)
\ee
and
\be
\int dp\,p^{2n-1}\,|\tilde{f}(p)|^2\,|\Phi(\l p)-\Phi(0)|^2
\longrightarrow 0
\ee
for $\l\downarrow 0$, showing the convergence of (\ref{f}).

                                                We thus obtained
for each $n$ and each $\Phi\in P_n H\cap\CC^\infty$ with
the complex number $\Phi(0)\neq 0$ a multiple of a unitary map
$
V_{n,\Phi}\;:\;L^2(\Rp,p^{2n-1}dp)\longrightarrow  P_n H
$
which is defined on the dense set $\{\tilde{f}|_{\rp}\,|\,f\in
\D(\R)\}$ by
\be
 V_{n,\Phi}\;:\;\tilde{f}|_{\rp}\longmapsto\Phi(0)\ |
                             (\tilde{f}|_{\rp})\!>\ :=\
\lim_{\lambda
        \downarrow 0}\l^{-n}\int da\,f(a)\;T(a)\,D(\l)\,\Phi
 \ee
 and intertwines the irreducible representations of $\tilde{G}$.

\subsection{Definition of Pointlike Localized Charged Field Operators}

Now, we come to the definition of pointlike localized fields.
Take a local observable
$A\in\A(I_0)$, $I_0\in\k.$
Let $\rho$ be a localized and transportable endomorphism of $\A$
inducing a charged sector with finite statistics,
                           let $e$ be a field bundle
                                        multi-index of type
$(0,\rho,\rho)$. Then $F=F(e,A)$ is a local element of the
reduced field bundle. We want $F$ to be regularized, i.e.\ we
choose $F$
                           such that $\tilde{g}
                                      \mapsto \alpha_{\tilde{g}}
                                                     (F)$
is $\CC^\infty$ in the strong operator topology.
                      Let now $n\in\Rp$
and $P_n$ the projection on the subspace of conformal dimension
$n$ in $H_{red}$.

Then
                                     the vector $P_n F\OM$
is $\CC^\infty$. Hence,
                        we may define operator
valued distributions $\varphi_{n,F}^I$ on $\D(I)$, $I\in\k$, with
a domain of definition dense in the vacuum Hilbert space by
\be
\label{b}
\varphi_{n,F}^I(f)\,B'\OM:=\pi_\rho(B')\,V_{n,P_n F\Omega}\,
\tilde{f}|_{\rp},\;\;\;
f\in\D(I),\;B'\in\A(I').
\ee

Let now $e$ be a field bundle
                                        multi-index of arbitrary
                                                       type
$(\alpha,\rho,\beta)$. Then $F=F(e,A)$ is a local element of the
reduced field bundle mapping $H_\alpha$, the copy of $H$ associated
with the superselection sector $\alpha$, onto $H_\beta$, the copy
of $H$ associated with the superselection sector $\beta$.
In the reduced field bundle one can find \cite{FRS2} the following
relations of an exchange algebra with structure constants $R$\,:
\be
F(e_2,A_2)\,F(e_1,A_1)=\sum_{f_1\circ f_2}R^{e_2\circ e_1}_{f_1
\circ f_2}(+/-)\,F(f_1,A_1)\,F(f_2,A_2)
\ee
whenever $F_1$ is localized in the right/left complement of the
localization domain of $F_2$.
Hence,                   we can define operator
valued distributions $\varphi_{n,F}^I$ on $\D(I)$, $I\in\k$, with
a domain of definition dense in the Hilbert space $H_\alpha$ by
\be
\varphi_{n,F(e,A)}^I(f)\,F(e',B')\OM:=\sum_{g'\circ g}R^{e\circ e'}
_{g'\circ g}(+/-)\,F(g',B')\,\varphi_{n,F(g,A)}^I(f)\,\OM\label{h}
\ee
for $f\in\D(I)$ and $F(e',B')\in\F_{red}(I')$
whenever it is localized in the right/left  complement of $I$
with respect to $\k$.
\subsection{Properties of the Charged Field Operators}

 First, we prove the closability of the operators
 $\varphi_{n,F}^I(f)$. We start with the case of field bundle
multi-index $e$ of type $(0,\rho,\rho)$.
\medskip

{\bf Theorem:} Let $n\in\N,\ I\in\k,\ f\in \D(I),\
B'\in\A(I')$, $G'=F(e,C')\in\F_{red}(I')$ and let $F=F(e,A)$ be a
                              regularized local element of the
 reduced field bundle.
With the linear operator reversal $F\mapsto\hat{F}$ and the
antilinear charge conjugation operation $F\mapsto\bar{F}$ both
                                                    defined in
\cite{FRS2} and with the statistical phase $k_\rho$
                                                 we then have
\beam
(G'\OM,\v_{n,F}^I(f)\,B'\OM)&=&(\frac{1}{k_\rho})^{(+/-)1}\,
                              (\hat{G'}\,\v_{n,\bar{F}}^I(\bar{f})\,
\OM,B'\OM)\\
&=&
(\v_{n,F^{\ast}}^I(\bar{f})\,G'\OM,\,B'\OM)\\
\v_{n,F}^I(f)^{\dagger}
                                                     &=&
\v_{n,F}^I(f)^{\ast}|_{{\cal F}_{red}(I')\Omega}\;=\;
\v_{n,F^{\ast}}^I(\bar{f})
\eeam
$\v_{n,F}^I(f)$ is closable because
$\v_{n,F}^I(f)^{\ast}$ has a dense domain.
\medskip

{\bf Remark:}
Here and in following proofs we heavily use the property of
``weak locality":

Let $F,G$ be two local elements of the reduced
field bundle, $F$ leading from the vacuum sector to a charged sector
$\rho$, $G$ leading back from $\rho$ to the vacuum sector.
In \cite{FRS2} has been proven that
\be
GF=(\frac{1}{k_\rho})^{(+/-)1}FG
\ee
whenever $F$ is localized in the left/right complement of $G$.
``Weak locality" is the reminiscent of the Haag-Kastler axiom
``locality" in the exchange algebra of the reduced field bundle
in low-dimensional quantum field theory.
\medskip

{\bf Proof:}
The Casimir operator associated to the representation $U(\cdot)$
of the                                 universal
 covering of the                                      Lie group
$\SL$
has the following spectral decomposition \cite{Lan}\,:
\be
C_{\tilde{G}}=\sum_{i\in\r}^{P_i\neq 0}
                                   i(i-1)\,P_i\,.\label{c}
\ee
$C_{\tilde{G}}$ is a second order differential
operator in $\tilde{G}$\,. Hence, it is a local operator
in contrast to the global projector
               $P_n\,$\,.

Some algebra leads to
\beam \lefteqn{%
(G'\OM,\v_{n,F}^I(f)\,B'\OM)}\nn\\
&=&\lim_{\lambda\downarrow 0}\int_{\r}dx\,f(x)\,(\pi_\rho(
                                                 B')^{\ast}G'\OM,U(x)\,
D(\l)\,\l^{-n}P_nF\OM)\nn\\
&=&\lim_{\lambda\downarrow 0}\int_{\r}dx\,f(x)\,(\pi_\rho(
                                                 B')^{\ast}G'\OM,U(x)\,
D(\l)\,\l^{-n}\left(\prod_{0<i<n}^{i\in {\bf Z}+n}
                                            \frac{C_{\tilde{
                                             G}}-i(i-1)}{n(n-1)-i(i-1)}
\right)P_nF\OM),\nn\\
&=&\lim_{\lambda\downarrow 0}\int_{\r}dx\,f(x)\,
\nn\\
&\;\;&
                                                (\pi_\rho(B')
                                                 ^{\ast}G'\OM,U(x)
\,D(\l)\,\l^{-n}\left(\prod_{0<i<n}
                             ^{i\in {\bf Z}+n}
                               \frac{C_{\tilde{
                                       G}}-i(i-1)}{n(n-1)-i(i-1)}
\right)(1-\sum_{0<i<n}
                ^{i\in {\bf Z}+n}
                  P_i)\,F\OM),\nn
\eeam
(Because of equation (\ref{c})
     the polynomial in $C_{\tilde{G}}$
                           has the property to act as the identity
operator on $P_n$ and as the zero operator on all $P_i, i<n$.
As a consequence of
the
       conformal cluster theorem,
                                 the contribution
of conformal energies $\geq n+1$ vanishes uniformally
                                          in the limit $\lambda
\rightarrow 0\,.$ Since the conformal rotation by $2\pi$ leaves the
observable algebra invariant, the conformal energy in a superselection
sector takes values in ${\bf Z}+n$ with a fixed conformal
 dimension $n$   )
\beam
&=&\lim_{\lambda\downarrow 0}\int_{\r}dx\,f(x)\,(\pi_\rho(B')
                                                ^{\ast}G'\OM,U(x)\,
D(\l)\,\l^{-n}\left(\prod_{0<i<n}
                           ^{i\in {\bf Z}+n}
                             \frac{C_{\tilde{
                                     G}}-i(i-1)}{n(n-1)-i(i-1)}
\right)F\OM)\nn\\
&&\;\;\;\mbox{we use weak locality}\nn\\
&=&\lim_{\lambda\downarrow 0}\int_{\r}dx\,f(x)\,(\frac{1}{k_\rho})^
{(+/-)1}\,
                                                (U(x)\,
D(\l)\,\l^{-n}\left(\prod_{0<i<n}
                           ^{i\in {\bf Z}+n}
                             \frac{C_{\tilde{
                                             G}}-i(i-1)}{n(n-1)-i(i-1)}
\right)\bar{F}\OM,\bar{G'}B'\OM)\nn\\
&=&(\frac{1}{k_\rho})^{(+/-)1}\,
   (\hat{G'}\,\v_{n,\bar{F}}^I(\bar{f})\OM,B'\OM)\nn\\
&=&
   (\v_{n,F^{\ast}}^I(\bar{f})\,G'\OM,B'\OM)
\eeam
with the definition of $\v_{n,F^\ast}^I$\,.

\medskip
Next, we argue that the closability theorem can be generalized to
fields with field bundle multi-index $e$ of arbitrary type $(\alpha,
\rho,\beta)$:
Since the reduced field bundle only considers superselection
sectors with finite statistics \cite{FRS2}, the definition (\ref{h}) of
charged fields of arbitrary type only contains a finite sum.
It is therefore a straight-forward calculation to show the
closability of charged field operators of arbitrary type.
Yet, we were not able to give an explicit expression for
the adjoint operator on a dense domain of definition, because
in the general case of operators with multi-index of arbitrary
type we have to use the full exchange algebra instead of weak
locality in the proof of the above theorem.

 Moreover, the closures of the charged field operators are affiliated
to the associated local von Neumann algebras of the reduced field
bundle: The commutant of the von Neumann algebra of
the reduced field bundle localized in $I\in \k$
                         is given by the
algebra of observables in $I'$
                                     represented
                                            on the Hilbert space
of the full theory $H_{red}$.
Therefore, the proof of
 Prop.\,2.5.9 in \cite{BrR} and more detailed in \cite{Joe1}
for neutral fields can be transferred to the case of charged fields.
For closed field operators $\v^I$ localized in $I\in \k$ we
obtain the affiliation relation
\be
\v^I\,\F_{red}(I)'\,\subseteq\,\F_{red}(I)'\, \v^I\,.
\ee

The existence of sufficiently many charged field operators
                                         such that their linear span
applied to the vacuum vector is dense in the Hilbert space $\H_{red}$
                                                           can be
proven exactly the same way
       as done in \cite{FrJ} for the vacuum case.
For each $P_nF\OM\neq 0$ a non-zero field can be constructed.
In a similiar way elements $F_i$ of the reduced field bundle
                              with a field bundle multi-index
$e$ of arbitrary type $(\alpha,\rho,\beta)$ can be chosen such that
the fields $\v_{n,F_i}$ are non-zero and orthogonal.

 Finally,
 one can easily see that the charged fields tranform covariantly:
\be
U(\tilde{g})
         \,\varphi_{n,F}^I(f)\,U(\tilde{g})
                                 ^{-1}=\varphi_{n,F}^{gI}(f_{
                                                  \tilde{g}}^{
 (n)})
 \ee
with $n\in\R$, $F$ localized in $I_0\in\k$ and regularized,
                       $\tilde{g}\in \tilde{G}$, with the
covering projection $\tilde{g}\mapsto g$,
$I,gI\in\k$ and $f\in D(I)$.

\medskip
\section{Consequences of the Existence of Pointlike Localized
Charged Fields}
The coexistence of the formulation of quantum field theory in terms
of Haag-Kastler nets of von Neumann algebras
                     and in terms of unbounded field operators with
pointlike localization can be used to derive important
                                             structural results
of the theory.
In the present paper, we
                     derive for all charged sectors with finite
 statistics                                         of
 a conformally invariant theory in 1+1 dimensions
                              a generalized Bisognano-Wichmann property,
the PCT theorem, the Spin-Statistics theorem, an equivalence theorem
for both formulations, additivity and an operator product expansion.
               In \cite{FrJ}, we had the coexistence of both
formulations in the vacuum sector and in the vacuum sector
the
Bisognano-Wichmann property, Haag duality, PCT covariance, equivalence
of both formulations, additivity and a operator product expansion
                                       could be derived.

\subsection{PCT, Spin\,\&\,Statistics, and All That}
We start with a proof of the Spin-Statistics theorem for conformally
invariant quantum field theory in 1+1 dimensions. We use the
argument of \cite{FRS2}.
\medskip

{\bf Spin-Statistics Theorem:}
The statistical phase $k_\rho$
and the chiral scaling dimension $n_\rho$ of a conformal field
$\v$ with arbitrary
          charge $\rho$ with finite statistics
                        fulfill the following
relation
\be
e^{2\pi in_\rho}=e^{2\pi in_{\bar{\rho}}}=k_\rho.
\ee

{\bf Proof:}
Let $F,G$  be regularized
           elements of the reduced field bundle localized in
disjoint intervals and $n\in\R$ such that $P_nF\OM\neq 0$\,.
With weak locality we obtain
\be
(\v_{n,F}(x)\,\OM,\,\v_{n,G}(y)\,\OM)=
k_\rho^{\mbox{{\footnotesize sign}}(y-x)}\,
(\v_{n,\bar{G}}(y)\,\OM,\,\v_{n,\bar{f}}(x)\,\OM)\,.
\ee
The explicit knowledge of the transformation properties of
the charged conformal two-point-function leads to
\be
(\v_{n,F}(x)\,\OM,\,\v_{n,G}(y)\,\OM)=
e^{2\pi in\,\mbox{{\footnotesize sign}}(y-x)}\,
(\v_{n,\bar{G}}(y)\,\OM,\,\v_{n,\bar{f}}(x)\,\OM)\,.
\ee
Since it has been shown in the last section that the charged fields
generate a dense set in $H_{red}$ from the vacuum sector $\OM\,$,
a comparison of the phases yields the theorem.

Next, we derive a generalization of the
Bisognano-Wichmann result to the reduced field bundle formalism.
The Tomita-Takesaki theory \cite{Tak,BrR}\
assigns to every pair of a von Neumann algebra \A\
and a cyclic and separating vector
$\Psi$ a closable, antilinear
operator:
\be
S_o:A\Psi\mapsto A^\ast\Psi\ \ \mb{for all}\ A\in\A\,.
\ee
$S$, the closure of $S_o$,
                    has a polar decomposition $\;S=J\triangle^{
\frac{1}{2}                                                    }\;$
and its components fulfill
a couple of relations:
\be
J=J^\ast,\ \ \ J^2={\bf 1},\ \ \ \triangle^{-\frac{1}{2}
                                            }=J\triangle^{
 \frac{1}{2}                                              }J,\
\ \ J\A J=\A'.      \ee
The set of operators
$\triangle^{it},\ t\in\R$, generate the group of modular
automorphisms:
\be
\triangle^{it}\A\,\triangle^{-it}=\A\,.
\ee

If $\Psi$ is chosen to be the vacuum vector $\OM$ and
\A\ an algebra of local observables in a conformally covariant
Haag-Kastler net in 1+1 dimensions, \cite{FrG,BGL} have
identified $J$ as the geometric reflection of the localization
domain onto its complement on the circle and the modular automorphism
group as the supgroup of conformal transformations which leave
the localization domain invariant.

On the Hilbert space $H_{red}$ of all charged sectors
with finite statistics
               we consider a generalized modular structure
based on the charge conjugation operation $F\mapsto\bar{F}$
instead of the operator adjoint $F\mapsto F^\ast\,$.
In the following, $S_I=J_I\Delta_I^{\frac{1}{2}}$ is defined as
the closure of the operator defined by the mapping
$F\OM\mapsto\bar{F}\OM$ with $F\in\F_{red}(I)$ for $I\in\k\,$.

In order to find a PCT-operator on $H_{red}$ we define the
antilinear Operator $\Theta\,:$
\be
\Theta\,\v_{n,F}(x)\,\OM:=(-1)^n\,
                          \v_{n,\bar{F}}(-x)\,\OM
\ee
for $n,x\in\R$ and $F\in\F_{red}$ regularized.
It can be easily seen that $\Theta$ is antiunitary and commutes
PCT-covariantly with the representation $U(\tilde{G})\,$.
To prove that $\Theta$ acts geometrically on the reduced field
bundle and on the charged field operators we first derive
the following generalization of the Bisognano-Wichmann result
to charged sectors with finite statistics.
\medskip

{\bf Bisognano-Wichmann Theorem:}
Let $k^{\frac{1}{2}}$ be the operator defined by its
eigenvalues $k_\rho^{\frac{1}{2}}$ on the Hilbert spaces
                                          $H_\rho$
associated with the different superselection sectors. Let
                                  $V(\cdot)$ be
the dilation subrepresentation of $U(\cdot)$ and let
$\Theta$ and $S_I$ be defined as above. We get as a
generalization of the result of Bisognano and Wichmann
\be
S_{\rp}=k^{\frac{1}{2}}\Theta V(i\pi)\,,\;\;\;S_{\rmm}=
k^{-\frac{1}{2}}\Theta V(-i\pi)\,.
\ee

{\bf Proof:}
Take the proof in \cite{Joe2} and use weak locality instead of
locality.
\medskip

{\bf Remark:}
With this result, we see a posteriori that we could have simplified
the construction of charged fields.
         Instead of using reducible modules $P_n H_{red}$ we
could have started with irreducible modules $P_\tau
H_{red}$. As shown in \cite{FrJ} for the vacuum sector,
the Bisognano-Wichmann result suffices to prove the closability
of field operators constructed with projectors $P_\tau$ on
irreducible modules.

\medskip

As a consequence of the identification of a generalized modular
structure with objects with well known geometrical meaning
in the Bisognano-Wichmann theorem above,
we are able to derive PCT covariance in the full theory
\medskip

{\bf PCT Theorem:}
$\Theta$ acts geometrically on the reduced field bundle
\be
\Theta \F_{red}(\Rp) \Theta =\F_{red}(\Rm)
\ee
                                                     and on the
charged field operators
 \be
 \Theta \v_{n,F}(x) \Theta =\v_{n,\Theta F\Theta}(-x)
 \ee
 with $n,x\in\R$ and $F\in\F_{red}$ regularized.
\medskip

{\bf Proof:}
The generalized modular structure based on the
antilinear chage conjugation operation $F\mapsto\bar{F}$
can be identified with the relative modular structure introduced
by Araki (see \cite{Iso,FRS2,Ara}).
   Then,             Araki's results on
relative modular operators imply the geometrical
action of $k^{-\frac{1}{2}}J_{\rp}=k^{\frac{1}{2}}
                                 J_{\rmm}=\Theta$ on
the reduced field bundle.
The geometrical action on the charged field operators then follows
by straight forward calculation.
\medskip

{\bf Remark:}
The PCT theorem can be obtained more directly. Another PCT operator
$\tilde{\Theta}$
can be constructed on the physical Hilbert space $H_{red}$ in
a natural manner: We already have a PCT operator on the
vacuum Hilbert space and we know how the PCT operation should
intertwine between the different copies of the vacuum Hilbert space.
$\tilde{\Theta}$, too, can be shown to act geometrically on the
reduced field bundle and on the charged field operators.

\subsection{Equivalence between the ``algebra picture" and the
``distribution picture"}

                                           The algebra generated by
                         polar and spectral decomposition
of all $\varphi_{\tau,e}
                       ^I(f)^{**}$, $\tau$ irreducible, $e$ of arbitrary
  type                                                 and $f\in \D(I),$
             is invariant under the generalized modular
automorphisms ${\rm Ad}\Delta_I^{it}$ introduced in subsection\,4.1 and
has the vacuum $\OM$ as a cyclic vector.
Hence, it coincides with $\F_{red}(I)$ \cite{Tak}.

Thereby, we proved the equivalence of the formulation in terms
of nets of von Neumann algebras
                      and in terms of unbounded field operators
with pointlike localization.
Without any loss of information
                             one can change between
                                            the ``algebraic
picture" and the ``distribution picture".

We now prove the
additivity of the                                  local
von Neumann algebras of the reduced field bundle:
                                   If
$I=\bigcup_\alpha I_\alpha$ with $I, I_\alpha \in\k$,
then
\be\F_{red}(I)=\bigvee_\alpha \F_{red}(I_\alpha )\ee
where
$\bigvee$ denotes the generated von Neumann algebra.

This follows from the fact that the local von Neumann algebras
can be constructed by linear combinations of the decomposition
parts of the unbounded field operators without building products
of field operators.

Additivity can also be proven directly. Using again the
identification of the generalized modular automorphism
group with the subgroup of dilations, one can transfer the
proof in \cite{FrJ} from the vacuum sector to
all charged sectors with finite statistics.

\subsection{Operator Product Expansion in Charged Sectors}

In the Haag-Kastler framework, the existence of an
operator product expansion might be formulated as the existence of
sufficiently many field operators such that their linear span
applied to the vacuum vector is dense in the Hilbert space.
                               Actually, in \cite{FrJ}
                                          a stronger result
with local coefficients and covariance w.r.t.\ the modular
$*$-operation $S$ has been derived in the vacuum sector.
Here, we prove an operator product expansion
                                          for arbitrary charged sectors
with finite statistics.

\medskip

{\bf Theorem:} Let $I\in\k$ and $F\in\F_{red}(I)$.
We then obtain a local expansion
\be
F\,=\,\sum_{\tau,\alpha}\varphi_{\tau,\alpha}^I(f_{\tau,\alpha,F})\,
\label{d}\ee
into a sum over all sectors $\alpha$ with finite statistics
                                     and all
                    irreducible subrepresentations $\tau$
of $U(\tilde{G})$
with
\be
\mb{supp}f_{\tau,\alpha,F}\subset I,
\ee
which converges on $\F_{red}(I')\OM$ $*$-strongly (cf. the definition in
\cite{BrR}).

\medskip

For a $F$ with a
                field bundle multi-index $e$ of type $(0,\rho,\rho)$ and
   for a irreducible $\tau$ the two
                                     simultaneous conditions
\be \label{e}
P_\tau F\OM=\varphi_{\tau}(f_{\tau,F})\,\OM\;\;\;\mb{and}\;\;\;
P_\tau \bar{F}\OM=\overline{
                   \varphi_{\tau}(f_{\tau,F})}\,\OM
\ee
                                                 together
fully determine the testfunction $f_{\tau,F}$:
\beam
f_{\tau,F}(x)&=&-i^{2n-1}\int_{-\infty}^{x}dy_1\int_{-\infty}^{y_1}dy_2
\cdot\!\cdot\!\cdot\int_{-\infty}^{y_{2n-2}}dy_{2n-1}\
(\,\OM,\,
[\,\varphi_{\tau}(y_{2n-1})^*,F\,
]^\wedge\,\OM\,) \nn\\
&=&\frac{1}{2\pi}\int_{-\infty}^{\infty}dp\,e^{ipx}\,\frac{
\int_{-\infty}^{\infty}dy\,e^{-ipy}\,
(\,\OM,\,
[\,\varphi_{\tau}(y)^*,F\,
]^\wedge\,\OM\,)}{p^{2n-1}}. \label{g}
\eeam
We used the abbrevation $
[\,F,G
]^\wedge:=FG-\hat{G}\hat{F}$  with the operator reversal $F\mapsto
\hat{F}$\,.

{\bf Proof:} The formula for $f_{\tau,F}$ in (\ref{g}) follows
from straight forward calculation since the two conditions (\ref{e})
                                                           determine
the positive and negative energy content of $f_{\tau,F}$ respectively.
It remains to be shown that
$f_{\tau,F}$ has support in $I$\,.

Because of the nontrivial phases occuring in the relation of weak
locality, the commutator function in the reduced field bundle
has no compact support.
Instead, we show that the convolution of the commutator function
with the Fourier transform of $p^{-(2n-[2n])}$
\be
(\,\OM,\,
[\,\varphi_{\tau}(x)^*,F\,
]^\wedge\,\OM\,)\ \otimes x^{2n-[2n]-1}
       \,:=\,\int_{-\infty}^\infty dy\,
(\,\OM,\,
[\,\varphi_{\tau}(y-x)^*,F\,
]^\wedge\,\OM\,)\ y^{2n-[2n]-1}
\ee
has support in $I$: \\
               Let $x\in I'$\,. Without restriction of generality we
          assume                $x$ to be on the right side of $I$.
After a transformation  of variables
                                 $y\mapsto -y$ in the second term of the
sum of the convoluted commutator,
                                 we obtain with the Spin-Statistics
theorem                                    the phase factor that occurs
in the relation of weak locality.
                          Now, we can interprete the convolution as a
line integral over the boundary of the upper complex half plane
                   ${\bf C}_+$   with an
                                 integrand analytic in
${\bf C}_+$\,. The contribution of the infinite half circle to
the line integral vanishes because of the conformal cluster theorem.
Now, again without restriction of generality, we assume $F$ to
be regularized.
        Applying weak locality,
 one can then see that
                                                   the integrand is
continuous at $y=0$ and
           on the whole
                  boundary of ${\bf C}_+$\,.
Hence, with the theorem of Cauchy from complex analysis the integral
vanishes and we have proven that the convoluted commutator function
has support in $I$\,.

This support property and
            conformal cluster theorem
specifies the Fourier transform $G(p)$
                                        of the commutator function
$(\,\OM,\,
[\,\varphi_{\tau}(x)^*,F\,
]^\wedge\,\OM\,)$
to be of the form $p^{2n-1}\,H(p)$, with an appropriate
analytic function $H(p)$.
Therefore, using the Paley-Wiener theorem (\cite{Tre}, theorem\,29.2)
                                                      we see that
           the support of $f_{\tau,F}(x)=\tilde{H}(x)$
                                       is included in the support
of the convoluted commutator function. Hence, it is included in $I$.

The local expansion (\ref{d})
then follows directly from the result above and
             the    rules for charged field operators
defined on arbitrary sectors $\alpha$ with finite statistics.
\medskip

{\bf Remark:}
As a by-product of the proof above we obtain a result on the
charged two-point-function in a conformally covariant
Haag-Kastler net in 1+1 dimensions. Let $F,G$ be any two local
elements of the reduced field bundle with disjoint domains of
localization. Let
$P_kF\OM=P_k\bar{F}\OM=0,\,k<n$. Then the
conformal cluster theorem implies that the
two-point-function
$(\,\OM,\,G\,U(x)\,F\OM\,)$
decreases as $x^{-2n}$. With the argument of the proof above
its Fourier transform
can be written as $\Theta(p)\,p^{2n-1}\,H(p)$ with an
appropriate analytic function $H(p)$.

\end{document}